\documentclass[aps,prl,showpacs]{revtex4}
\usepackage{amssymb}
\usepackage{amsmath}

\usepackage{graphicx}
\usepackage{bm}

\begin{document}

\title{Quasi exact solution of the Rabi Hamiltonian}
\date{\today}
\author{Ramazan Ko\c{c}}
\email{koc@gantep.edu.tr}
\affiliation{Department of Physics, Faculty of Engineering University of Gaziantep, 27310
Gaziantep, Turkey}
\author{Mehmet Koca}
\email{kocam@squ.edu.om}
\affiliation{Department of Physics, College of Science, Sultan Qaboos University, PO Box
36, Al-Khod 123, Muscat, Sultanate of Oman}
\author{Hayriye T\"{u}t\"{u}nc\"{u}ler}
\email{tutunculer@gantep.edu.tr}
\affiliation{Department of Physics, Faculty of Engineering University of Gaziantep, 27310
Gaziantep, Turkey}

\begin{abstract}
A method is suggested to obtain the quasi exact solution of the Rabi
Hamiltonian. It is conceptually simple and can be easily extended to other
systems. The analytical expressions are obtained for eigenstates and
eigenvalues in terms of orthogonal polynomials.
\end{abstract}

\maketitle


Considerable attentions have been paid over the years to the solution of the
Rabi and Jahn-Teller(JT) Hamiltonians\cite{judd1, judd2, loor}. The $%
E\otimes \epsilon $ JT problems has been solved by Judd when certain
relations between the parameters of the Hamiltonian invoked\cite{judd2}.
Such solutions are known as Juddian isolated exact solutions. The problem
has been studied in the Bargmann-Fock space by Reik \textit{et al}\cite%
{reik1} and its canonical form has been obtained by Szopa \textit{et al}\cite%
{szopa1}. It has been proven \cite{reik2, szopa2} that the Rabi Hamiltonian
i.e. $E\otimes \beta $ JT system and $E\otimes \epsilon $ JT system are
mathematically identical.

In this letter we take a new look at the solution of Rabi Hamiltonian
through the method of quasi exact solvability. \ The $E\times \beta $
Jahn-Teller system coupled to a system executing harmonic oscillation which
is separated in energy by $2\mu $ is characterized by the Rabi Hamiltonian%
\cite{reik2}:%
\begin{equation}
H=a^{+}a+\kappa \sigma _{3}(a^{+}+a)+\mu (\sigma ^{+}+\sigma ^{-})
\label{eq:1}
\end{equation}%
where $\sigma ^{\pm }=\frac{1}{2}\left( \sigma _{1}\pm i\sigma _{2}\right) $
and $\sigma _{1},\sigma _{2},\sigma _{3}$ are Pauli matrices and the
parameter $\kappa $ is linear coupling constant. The Hamiltonian (\ref{eq:1}%
) can be expressed as a differential equation in the Bargmann-Fock space by
using the realizations of the bosonic operators, 
\begin{equation}
a^{+}=z,\quad a=d/dz.  \label{eq:2}
\end{equation}%
\ Substituting of (\ref{eq:2}) into (\ref{eq:1})we obtain a system of two
linear differential equation for the functions $\psi _{1}(z)$ and $\psi
_{2}(z):$ 
\begin{subequations}
\begin{eqnarray}
(z+\kappa )\frac{d\psi _{1}(z)}{dz}+(\kappa z-E)\psi _{1}(z)+\mu \psi
_{2}(z) &=&0  \label{eq:1a} \\
(z-\kappa )\frac{d\psi _{2}(z)}{dz}-(\kappa z+E)\psi _{2}(z)+\mu \psi
_{1}(z) &=&0  \label{eq:1b}
\end{eqnarray}%
where $E$ is the eigenvalue of the Rabi Hamiltonian. We eliminate $\psi
_{2}(x)$ between two equations and substituting 
\end{subequations}
\begin{equation}
z=\kappa (2x-1),\quad \psi _{1}(z)=e^{-2\kappa ^{2}x}\Re (x)
\end{equation}%
we obtain a second order differential equation%
\begin{eqnarray}
x(1-x)\frac{d^{2}\Re (x)}{dx^{2}}+\left[ \kappa ^{2}(4x^{2}-2x-1)+E(2x-1)-x+1%
\right] \frac{d\Re (x)}{dx} &&  \nonumber \\
+\left[ \kappa ^{4}(-4x+3)-E^{2}+2E\kappa ^{2}(-2x+1)+\mu ^{2}\right] \Re
(x) &=&0.  \label{eq:3}
\end{eqnarray}%
In order to solve (\ref{eq:3}) we first introduce the following linear and
bilinear combinations of the operators of the $sl(2,R)$ Lie algebra,%
\begin{equation}
J_{+}J_{-}+J_{-}J_{0}-jJ_{-}-4\kappa ^{2}J_{+}+(4\kappa
^{2}+2j-1)J_{0}+(j(4\kappa ^{2}-1)+\mu ^{2}-2j)=0  \label{eq:4}
\end{equation}%
which is quasi exactly solvable(QES)\cite{turb, koc}. The differential
realizations of the generators of the algebra is given by, 
\begin{equation}
J_{-}=\frac{d}{dx},\quad J_{0}=x\frac{d}{dx}-j,\quad J_{+}=-x^{2}\frac{d}{dx}%
+2j.  \label{eq:5}
\end{equation}%
The insertion of (\ref{eq:5}) into (\ref{eq:4}) leads to the following
differential equation,%
\begin{eqnarray}
x(1-x)\frac{d^{2}\Re (x)}{dx^{2}}+\left[ 2j(2x-1)+(x-1)(4\kappa ^{2}x-1)%
\right] \frac{d\Re (x)}{dx} &&  \nonumber \\
+\left[ 8j\kappa ^{2}(1-x)+\mu ^{2}-4j^{2}\right] \Re (x) &=&0.  \label{eq:6}
\end{eqnarray}%
The function $\Re (x)$ is a polynomial of degree $2j.$ The equations (\ref%
{eq:3}) and (\ref{eq:6}) are identical under the condition%
\begin{equation}
E=2j-\kappa ^{2}.  \label{eq:7}
\end{equation}%
The resulting differential equation(\ref{eq:6}) \ and the equation which we
have discussed in a paper is identical, if some parameters are reordered\cite%
{koc}. Now we can easily obtain the results given in the paper\cite{koc} by
defining the parameters%
\begin{eqnarray}
\alpha &=&\frac{1}{2},\quad \lambda =-4j(2\kappa ^{2}-j)-\mu ^{2},\quad
L=-2j-\frac{1}{2},\quad A=-S-\frac{1}{2}  \nonumber \\
q &=&\frac{16\kappa ^{2}}{(2S+1)^{2}},\quad S=\left[ 4j(j+1)+(4\kappa
^{2}+1)^{2}\right] ^{1/2}.  \label{eq:8}
\end{eqnarray}%
Then $\Re (x)$ can be expressed in terms of the the polynomial $P_{m}(\kappa
)$: 
\begin{equation}
\Re (x)=\sum\limits_{m=0}^{2j}\frac{4^{4j+m}j\Gamma (1-4j)\Gamma (2j)\sin %
\left[ (2j-m)\pi \right] P_{m}(\kappa )(-\kappa ^{2}x)^{m}}{\sqrt{\pi }%
(2j-m)\Gamma (m+1)}.  \label{eq:9}
\end{equation}%
Here $P_{m}(\kappa )$ satisfies the recurrence relation%
\begin{eqnarray}
4\kappa ^{2}(m-2j)P_{m+1}(\kappa )-\left[ (2j-m)(2j-4\kappa ^{2}-m)+\mu ^{2}%
\right] P_{m}(\kappa ) &&  \nonumber \\
+4\kappa ^{2}(m-2j)P_{m-1}(\kappa ) &=&0
\end{eqnarray}%
with the normalization $P_{0}(\kappa )=1.$ Certain properties of the
polynomial $P_{m}(\kappa )$ have been discussed in some recent works\cite%
{finkel}. The polynomial $P_{m}(\kappa )$ vanishes for $m\eqslantgtr 2j+1$
and the roots of \ $P_{2j+1}(\kappa )$ leads to the relations between the
parameters of the Hamiltonian. The first three of them are given by 
\begin{align}
P_{1}(\kappa )& =4\kappa ^{2}+\mu ^{2}-1  \nonumber \\
P_{2}(\kappa )& =32\kappa ^{4}+4(3\mu ^{2}-8)\kappa ^{2}+\mu ^{2}(\mu
^{2}-5)+4  \nonumber \\
P_{3}(\kappa )& =384\kappa ^{6}+16(11\mu ^{2}-54)\kappa ^{4}+8(3\mu
^{4}-29\mu ^{2}+54)\kappa ^{2}+\mu ^{2}(\mu ^{2}-7)^{2}-36.  \label{eq:11}
\end{align}%
These relations are exactly the same results obtained by the method of
Juddian isolated exact solution\cite{loor, kus}. The solution obtained for
the eigenfunction $\psi _{1}(x)$ can be substituted in (\ref{eq:1a}) to
determine the \ other component $\psi _{2}(x)$ of the wave function.

In conclusion we have shown there exists a quasi exact solution of the Rabi
Hamiltonian implying that $E\otimes \epsilon $ JT system also has a quasi
exact solution. The method given here can be extended other JT or
multi-dimensional atomic system problems. Another interesting implication of
the method is that the existence of the relation between the QES P\"{o}%
schl-Teller family potentials and Rabi systems. Details of the work is under
investigation.


\begin{thebibliography}{99}
\bibitem{judd1} Jamila S, Dunn L J and Bates C A 1993 \textit{J. Phys:
Condens. Matter }\textbf{5} 1493-1504; 2000\ Bosnick K A \textit{Chem. Phys.
Lett.} \textbf{317} 524-528

\bibitem{judd2} Judd B R 1979 \textit{J. Phys C: Solid State Phys.} \textbf{%
12} 1685-92

\bibitem{loor} Loorits V 1983 \textit{J. Phys C: Solid State Phys.} \textbf{%
16} L711-L715

\bibitem{reik1} Reik H G 1993 J. Phys. A: Math. Gen., \textbf{26} 6549

\bibitem{szopa1} Szopa M and Ceulemans A 1997 J. Phys. A: Math. Gen., 
\textbf{30 }1295

\bibitem{reik2} Reik H G and Wolf G 1994 J. Phys. A: Math. Gen., \textbf{27}
6907

\bibitem{szopa2} Szopa M and Ceulemans A 1996 \textit{J. Math Phys.} \textbf{%
37} 5402

\bibitem{kus} Kus M and Lewenstein M 1986 J. Phys. A: Math. Gen.,\textbf{19}
305

\bibitem{turb} Turbiner A V and Ushveridze A G 1987 \textit{Phys. Lett. }%
A126 181-3; Gonzalez-Lopez A, Kamran N, and Olver P J 1993 \textit{\ Commun.
Math. Phys}. 153 117-46

\bibitem{koc} Ko\c{c} R, Koca M, T\"{u}t\"{u}nc\"{u}ler H, "A Unified
Treatment of Quasi-Exactly Solvable Potentials I" submitted to \textit{J.
Phys. A: Math. Gen }Ref: A/136189/PAP

\bibitem{finkel} Finkel F, Gonzalez-Lopez A and Rodiriguez M A 1996 \textit{%
\ J. Math. Phys.} 37 3954-3972; Krajewska A, Ushveridze A and Walczak Z 1997 
\textit{Mod. Phys. Lett}. A\textbf{12} 1131-44
\end{thebibliography}
\end{document}